\documentclass[conference]{IEEEtran}
\IEEEoverridecommandlockouts
\usepackage{cite}
\usepackage{amsmath,amssymb,amsfonts}
\usepackage{algorithmic}
\usepackage{graphicx}
\usepackage{textcomp}
\usepackage{xcolor}
\usepackage[normalem]{ulem}
\usepackage[colorinlistoftodos,prependcaption,textsize=scriptsize]{todonotes}

\def\BibTeX{{\rm B\kern-.05em{\sc i\kern-.025em b}\kern-.08em
    T\kern-.1667em\lower.7ex\hbox{E}\kern-.125emX}}

\IEEEoverridecommandlockouts
\IEEEpubid{\makebox[\columnwidth]{978-1-6654-0627-7/22/\$31.00 ©2022 IEEE\hfill}
\hspace{\columnsep}\makebox[\columnwidth]{ }}
\begin{document}
\title{
Assessing Intel's Memory Bandwidth Allocation for resource limitation in real-time systems
}

\author{\IEEEauthorblockN{Giorgio Farina\textsuperscript{1}, Gautam Gala\textsuperscript{2}, Marcello Cinque\textsuperscript{1}, Gerhard Fohler\textsuperscript{2}}
\IEEEauthorblockA{(1) Università degli Studi di Napoli Federico II, Italy \\
(2) Technische Universit\"{a}t Kaiserslautern, Germany\\
\{giorgio.farina,macinque\}@unina.it \{gala,fohler\}@eit.uni-kl.de}
}

\maketitle

\begin{abstract}
Industries are recently considering the adoption of cloud computing for hosting safety critical applications. However, the use of multicore processors usually adopted in the cloud introduces temporal anomalies due to contention for  shared resources, such as the memory subsystem. In this paper we explore the potential of Intel's Memory Bandwidth Allocation (MBA) technology, available on Xeon Scalable processors. By adopting a systematic measurement approach on real hardware, we assess the indirect memory bandwidth limitation achievable by applying MBA delays, showing that only given delay values (namely 70, 80 and 90) are effective in our setting. We also test the derived bandwidth assured to a hypothetical critical core when interfering cores (e.g., generating a concurrent memory access workload) are present on the same machine.
Our results can support designers  by providing understanding of impact of the shared memory to enable predictable progress of safety critical applications in cloud environments.
\end{abstract}


\section{Introduction}


As seen in Ericson's Computing Fabric  \cite{ericsson}, and railway use case of EU projects such as SECREDAS \cite{gala:railway}, industries are exploring cloud computing and virtualization for hosting safety-critical applications \cite{CINQUE2022315} as clouds support ease of re-usability, maintainability, and reconfiguration while providing workload elasticity and higher availability. Thus, clouds help industries in reducing their running costs and carbon footprint. 
Industries can execute existing safety-critical applications without any significant modifications as Virtual Machines (VMs) in a cloud computing environment. These safety-critical VMs must run alongside other non-critical/best-effort VMs executing in the cloud. All VMs are unaware that they are running in a virtualized environment or know the actual underlying cloud nodes or hardware resources, thanks to the virtualization (abstraction) provided by the hypervisor.

Cloud nodes are based on COTS multicore processors, which are rich in shared resources to allow fast communication and improve the average resource utilization. Nevertheless, sharing introduces new forms of non-determinism and dependencies (in space and time) between parallel flows complicating the analysis of the Worst-Case Execution Time (WCET), which is fundamental for safety-critical applications. 

Recent research on real-time hypervisors focused mainly on CPU virtualization, adopting either hierarchical scheduling \cite{RT-XEN_xi2014real} or CPU partitioning \cite{jailhouse}, and implementing cache coloring \cite{modica2018supporting} to reduce inter-core cache interference in multicores. However, open problems regarding temporal contention (and consequent unexpected delays) are still observed when tasks/VMs co-executing on different cores need to access the memory (controller) simultaneously. \newline
Memory access contention deserves special attention for two main reasons: (i) memory accesses are unavoidable, and, contrarily to other sources of temporal contention\cite{Lundqvist1999TimingAI}\cite{Reineke2006ADA}, (ii) memory access interference can increase the WCET by several orders of magnitude, particularly when increasing the number of co-accessing cores \cite{Nowotsch2013, Pellizzoni2010WorstCD}. The picture is even more complex on modern multicore architectures for cloud computing where many cores (or hardware threads) share additional resources such as a Last Level Cache (LLC). For instance, the spatial contention in the LLC has a significant impact on the memory access contention as LLC writebacks lead to asynchronous memory accesses.
While memory accesses are synchronous with respect to the requester demand, asynchronous memory accesses (e.g., due to LLC writebacks) may depend on the memory requests from other cores, as the LLC is shared.

Existing work proposes several memory access regulation methods to limit the maximum interference for the worst case memory access latency analysis.
The basic idea is to limit the memory accesses from individual cores for each regulation period \cite{Yun2013MemGuardMB} \cite{Modica2018SupportingTA}. Thus, it is possible to distribute the available memory access bandwidth and obtain a tighter worst case memory access latency bound.
Software-based resource limitation approaches, such as Memguard \cite{Yun2013MemGuardMB}, require monitoring the memory requests (or cache misses) from each core and applying a throttling policy to prevent memory accesses from cores that have depleted the assigned memory budget for that regulation period.
Despite the design simplicity of software-based approaches, it is not possible to achieve high resolution and efficiency due to the coarse-grained regulation period duration and high overheads. In addition, the throttling policy of the approaches involve stalling the core execution to prevent memory accesses. Such a throttling policy is not suitable for cloud VMs as a core cannot continue its execution within the private context, such as private caches and functional units, reducing the private resources utilization.

\begin{figure} 	
	\centering
	\includegraphics[width=0.47\textwidth]{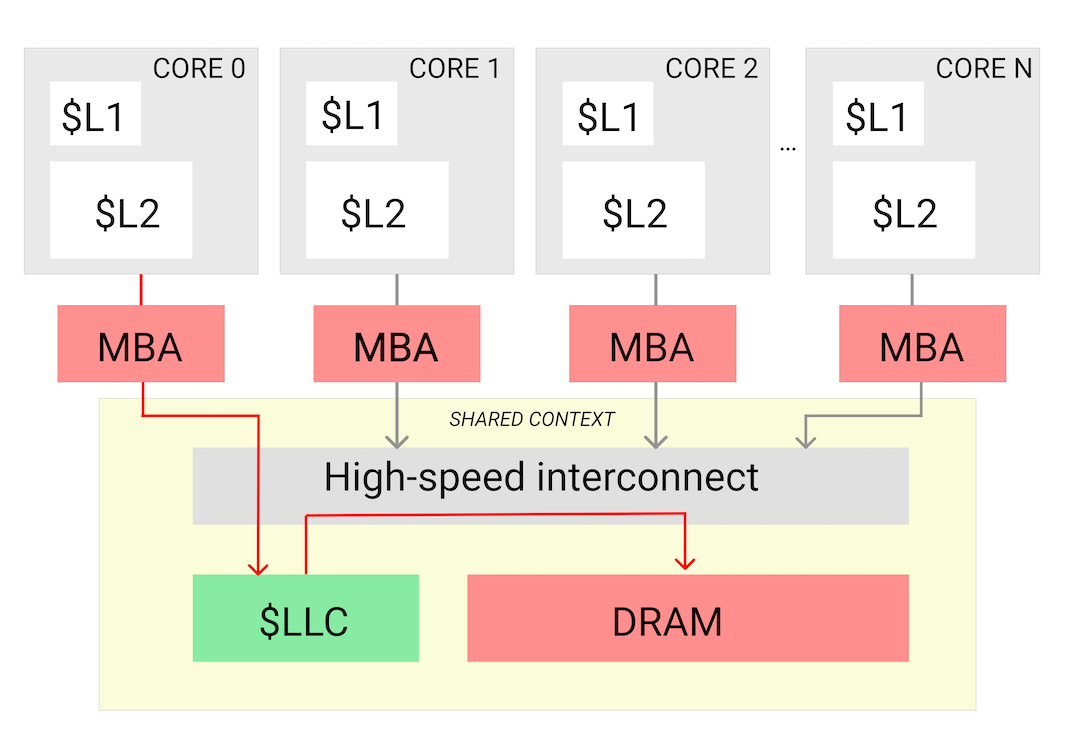}
	\caption{Abstract model of Intel Xeon memory hierarchy depicting the MBA controller between the cores and high-speed interconnect \label{fig:architecture}}
\end{figure}

A possible way to overcome the limitations of software-based approaches is to rely on 
hardware support. In this respect, Intel recently proposed a new hardware feature known as Memory bandwidth allocation (MBA) present in Xeon Scalable processors \cite{IntelRefManXeonScal}, depicted in Fig. \ref{fig:architecture}. MBA delays the requests going to the interconnect from a core's private context. Intel provides nine values between $10$ to $90$ in increments of $10$ representative of the delays that are inserted between the requests.
MBA can be used to apply indirect memory bandwidth throttling overcoming the limitations of software base-approaches as: (1) we have a fine-grained regulation period because the delays are inserted by hardware between requests (2) a hardware based method does not require software intervention, except for the initial setup,  to partition memory bandwidth statically  (3) the core execution is not stalled to prevent memory access, thus preserving the private resource utilization.

Despite these promising features, it is still unclear if and how MBA can be used for memory regulation to provide real-time guarantees to safety-critical VMs. As a matter of fact, the MBA controller  does not directly allocate memory bandwidth to cores, and hence the memory bandwidth allocation/regulation is an indirect consequence of the introduced delay.

In this paper, we present the results of a measurement study performed on Intel Xeon Scalable processors in order to derive the indirect memory bandwidth achieved by the cores through MBA for providing real-time guarantees. 
The main contributions and results of the study are:
\begin{enumerate}
    \item We propose a systematic approach to derive the indirect memory bandwidth limitation from delay values using a single-core analysis where we highlight the importance to consider also L2 writebacks within the design of experiments;
    \item We experimentally prove that only the 70, 80 and 90 MBA delay values are effective in our setting at limiting the memory bandwidth even in the presence of LLC write-backs (asynchronous memory accesses);
    \item We test the derived assured memory fetching latency for a core allocated to a critical VM when MBA limits the bandwidth of three interfering cores with a multicore analysis, showing the potential and effectiveness of the approach.
\end{enumerate}
The rest of the paper is structured as follows. Section \ref{sec:back} presents the related work, while the background on new Intel MBA feature is provided in Section \ref{sec:intel}. Our design of experiments is presented in Section \ref{sec:doe}, while Sections \ref{sec:one-core-res} and \ref{sec:multi-core-res} presents the results of single-core and multicore analysis, respectively. Finally Section \ref{sec:concl} ends the paper with final remarks and hints for future work.

\section{Related Work}
\label{sec:back}

The management of memory access time interference can be broadly classified in two approaches (extending the classification in \cite{Kotaba2013MulticoreIR}): (i) interference estimation, to take it into account in worst case execution time analysis, and (ii) interference regulation by either enforcing execution models or limiting the access to the memory controller at runtime. In the following we review the literature falling into these two categories.

\subsection{Interference Estimation Approaches\label{sec:anal_interf}}
The idea is to analytically estimate the maximum number of conflicts through static analysis. Under a worst-case perspective, the first goal is to classify accesses depending on the addressed hierarchical memory level. Such analysis gets complicated with shared caches. In \cite{Liang2009TimingAO} each access is first analyzed in the absence of inter-core cache conflicts, and subsequently, a separate inter-core cache conflict analysis refines each classified access. \cite{Chattopadhyay2013ScalableAP} improves the estimation of inter-core cache conflicts combining abstract interpretation and path-sensitive verification (e.g., model checking and symbolic execution). In \cite{Chattopadhyay2010ModelingSC}, \cite{Kelter2011BusAwareMW}, \cite{Chattopadhyay2012AUW} shared cache and bus analysis are combined, applying a time-division bus arbitration. \newline
Estimation approaches suffer from several drawbacks. 
First, the computational complexity increases with the number of tasks.
Second, the arbiter is supposed to be deterministic, enabling time-division multiple access (TDMA). However, this is not possible on modern COTS due to their optimization policies. Third, there are no independent partitions: the estimation validity is stictly coupled with the tasks under analysis.


\subsection{Interference Regulation Approaches \label{sec:regulate_approaches}}
\subsubsection{Approaches based on execution models}
The basic idea of execution models is to split real-time tasks into different phases and then enforce a high-level co-scheduling mechanism. 
The co-scheduler rules implicitly define the contention degree without relying on low-level arbiters.
The memory accesses are then serialized in several phases, and these phases can be overlapped by the co-scheduler achieving a predictable degree of interference. 
In \cite{Pellizzoni2011APE}, \cite{Boniol2012DeterministicEM} the co-scheduler implements a TDMA policy. In particular, 
the work in \cite{Pellizzoni2011APE} proposes PREM handling the contention of main memory between a single-core and several peripherals. Whereas, the work in \cite{Boniol2012DeterministicEM} extends the concept of the execution model to a multicore platform dividing the execution phase into communication and local execution slices. \newline
Execution models are invasive, as they require program or compiler modifications. Moreover, the TDMA policy is not scalable with the number of cores. 

\subsubsection{Approaches based on resource limitation}
The aim of these approaches is to make the maximum interference  predictable and deterministc, by limiting the processing element accesses, with no modification to the original program. These solutions are usually based on a throttling policy to stop the processing element access in case of excessive request, and a monitoring policy for counting the issued requests. \newline
Notably, \cite{Bellosa1999TheTD} are among the first ones to perceive the potentiality of performance counters to monitor shared resource accesses. The idea is to capture on each processing element (PE) the only interesting available events, such as last-level cache hits and references to extract the number of LLC misses (representing synchronous accesses). Nowadays, LLC misses are supported as non-architectural events on modern COTS and are also used to monitor PE accesses, \cite{Yun2012MemoryAC, Yun2013MemGuardMB}. \newline
As throttling policy, \cite{Bellosa1997ProcessCC} inserts idle-loops inside TLB-miss handler, while the works in \cite{Yun2012MemoryAC, Yun2013MemGuardMB} propose to deschedule the task.
The work in \cite{Nowotsch2013} starts from the computation of the \textit{interference-sentitive} WCET taking into account a safety bound for shared-resource interference when at runtime resource capacity enforcement is used to limit the actual resource usage of each process, which are executed following a round robin scheduling scheme. Such pessimistic model is based on a strong assumption on the interference delays. \newline
As software-based resource limitation approaches present the already mentioned efficiency and resolution issues, in the rest of the paper we focus on the new capabilities offered by Intel's MBA technology.


\section{New Intel Capabilities}
\label{sec:intel}
In this work we focus on the Intel Cascade-Lake micro-architecture, which inherited from the Skylake-Server micro-architecture an exclusive LLC and a mesh as a network of interconnect. An LLC exclusive frees the last level cache to be as large as the sum of all core private caches, saving the private cache size scalability (i.e., we went from 254KB to our 1MB L2 cache size). At the same time, the mesh allows higher parallelism in the memory access and last level cache.  \newline
It is important to take into account the exclusivity property of the LLC when designing the workload for the measurements (see next section). An exclusive LLC does not include all private cache lines, improving the overall cache space utilization. Such property influences the handling in response to an LLC miss where the new fetched cache line is not brought in LLC to keep the inclusivity.\newline
The considered micro-architecture, as well as all the new Intel Xeon processors, supports temporal and spatial isolation at the hardware level through Resource Director Technology (RDT) \cite{IntelRefMan}. We can associate spatial and temporal properties with a class of service (CLOS) and then bind the CLOS to a virtual core. \newline
One physical core cannot change the class of service of another core. However, it can configure the properties of any CLOS. \newline
RDT makes available monitoring and allocation features. \newline
RDT allocation features enable inter-core resource partitioning.
While the Cache Allocation Technology (CAT) and Code and Data Prioritization (CDP) deal with the LLC spatial partitioning, the Memory Bandwidth Allocation (MBA) addresses the temporal memory contention. The MBA controls a configurable delay for requests going to the network of interconnect from the core private context. By configuring such delays, we can apply an indirect limitation in memory access without impacting the use of private resources such as private caches and functional units. Operationally, a MBA delay value can be associate with a class of service (CLOS), e.g., to limit the memory bandwidth of cores hosting non-critical VMs. If two virtual cores of the same physical core have different delays, the larger delay will override the other.
In our case, disabling the Hyper Threading (HP) in the experiments, we have a side effect \cite{IntelRefManXeonScal}; in particular, the non-active hardware threads are assigned to the CLOS0 by default even though they are disabled. Hence, the CLOS0 has not been used. Otherwise, the hidden delay value of the non-active virtual core (the delay value associated with the CLOS0) overrides the MBA delay value of the active virtual machine on the same physical core when the first is higher than the second one.

\section{Design of Experiments}
\label{sec:doe}
We design a set of experiments to perform two main analyses.
The single-core analysis is based on a systematic approach to deriving the indirect memory bandwidth limitation achieved when the MBA delays the requests. 
Contrarily, the multi-core analysis is focused on the guarantees rather than the limitation.
Multi-core analysis tests the memory bandwidth achieved by a core executing a critical VM when the MBA (indirectly) limits memory accesses from 3 interfering cores (performing best-effort tasks).
Both, single-core and multi-core analysis, share the following architectural parameters (Table \ref{tab:share_parameters}). We use only one socket (one processor) composed of eight cores at most. We set the number of active cores via BIOS configuration. We only used one memory channel in both analyses. 
We set the DRAM refresh rate to periodic and the CPU core frequency to maximum (no power saving) in the BIOS. These parameters helped keep the latency variability to a minimum and maximize the memory requests rate. Similarly, we disabled the Pre-fetcher and Hyperthreading to reduce variability since hardware threads on the same physical core share the same “queue” to the memory, while the prefetcher can preload several lines after a single cache miss. 
Experiments are conducted on an Intel Xeon Silver processor, Cascade-Lake microarchitecture, with 8 cores, 2,10 GHz, 512KB L1 cache (256KB out of 512 KB are dedicated for the Code), 1MB L2 cache, and 11MB LLC. 
Next subsections present the design of experiments for both analyses. Our measures achieve \sout{the} 99\% of accuracy calculated as the 99\% mean confidence interval using a student's t distribution divided by the sample mean.
\begin{table}
\caption{Design of experiments: shared parameters \label{tab:share_parameters}}
\begin{center}
\begin{tabular}{ |c| c| }
\hline 
\textbf{Shared parameters} & \textbf{setting}
\tabularnewline
\hline 
\hline 
Sockets no. & 1
\tabularnewline
\hline 
Used DRAM channel & 1
\tabularnewline
\hline 
Prefetcher & disabled 
\tabularnewline
\hline 
SMT &disabled
\tabularnewline
\hline
DRAM refresh cycles & periodic
\tabularnewline
\hline
Core frequency &Maximum
\tabularnewline
\hline
\end{tabular}
\end{center}
\end{table}

\subsection{Single-core analysis}
Single-core analysis has the objective to estimate the memory access limit for an interfering core when MBA is enabled with different delays. We exclude 20 and 30 delays from the analysis as they are erroneous on our processor (as reported in an Intel errata \cite{IntelRefManXeonScal}) and produce the same results as delay value 10.
Table \ref{tab:doe_onecoreanalysis} describes the design of experiments of the single-core analysis. Our experiments measure the latency execution variability for different kinds of workloads: memory read workload considers only synchronous memory accesses, while memory write workload accounts also the asynchronous memory writes. One of the main contributions is to specialize the memory read workload regarding the L2 writebacks as motivated later.  The number of observations achieves more than 99\% of accuracy.

\begin{table}
\caption{Design of experiments: single-core analysis \label{tab:doe_onecoreanalysis}}
\begin{center}
\begin{tabular}{ |c| c| }
\hline 
\textbf{Parameters} & \textbf{setting}
\tabularnewline
\hline 
Cores no. & 1 
\tabularnewline
\hline 
\hline 
\textbf{Factors} & \textbf{values}
\tabularnewline
\hline 
MBA delay& $d\in D, D=\{10; 40; 50; 60; 70; 80; 90\}$
\tabularnewline
\hline
Workload &$k \in K, K=\{R^{I}; R^{II}; W^{II}\}$\\& \begin{tabular}{ c c c }
$R^{I}:$ \{&$R^{II}$: \{ &$W^{II}$: \{ \\8192  CAS R,&8192 CAS  R,&8192  CAS R,
\\no  L2WBs,&with  L2WBs,&with  L2WBs,
\\no LLCWBs\}&no LLCWBs\}&with LLCWBs \}
\end{tabular}
\tabularnewline
\hline 
\hline 
\textbf{Metric} & \textbf{description}
\tabularnewline
\hline 
Memory & memory fetching latency in terms of CPU cycles\\fetching&
$L_{k;d}$, $k \in K, d \in D$ \\latency &
\tabularnewline
\hline
\end{tabular}
\end{center}
\end{table}

\subsubsection{Memory read workload}
A read request in the worst case corresponds to a memory read operation when there is an LLC miss. Since we consider this worst-case in our workload, read request and memory read operation are interchangeable terms.\newline
However, we identify two different L2 traffic patterns for this worst-case due to the LLC exclusivity in the processor. \newline
In the exclusive architecture, L2 writebacks (L2WBs) to the LLC happen whenever new memory is read and the L2 set is full, even though the evicted L2 cache line is clean (contrarily to the inclusive architecture). 

\begin{figure}[t!]
\centering
\includegraphics[width=.48\textwidth]{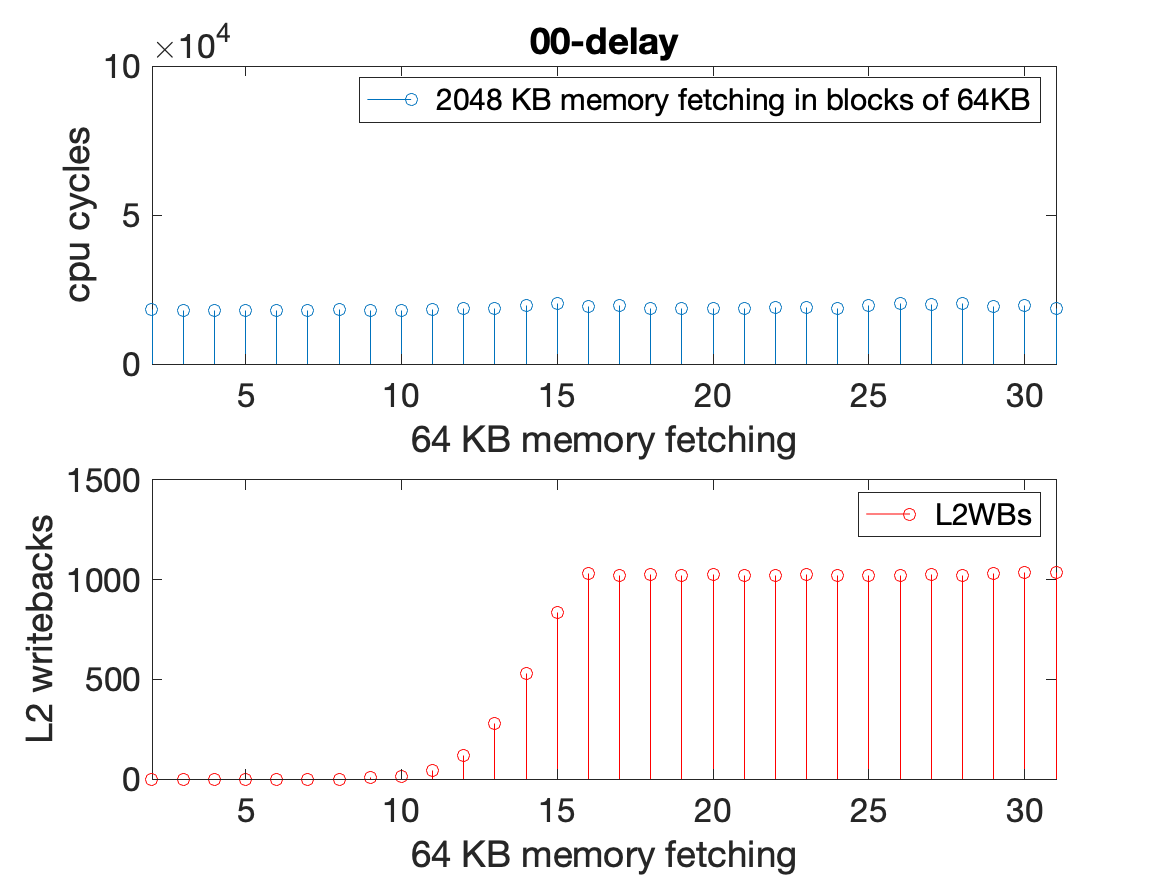} \
\caption{2048 KB memory fetching in several blocks of 64KB when the MBA is disabled (0-delay)\label{fig:fineg_l2_wb_00}}
\end{figure}
\begin{figure}[t!]
\centering
\includegraphics[width=.48\textwidth]{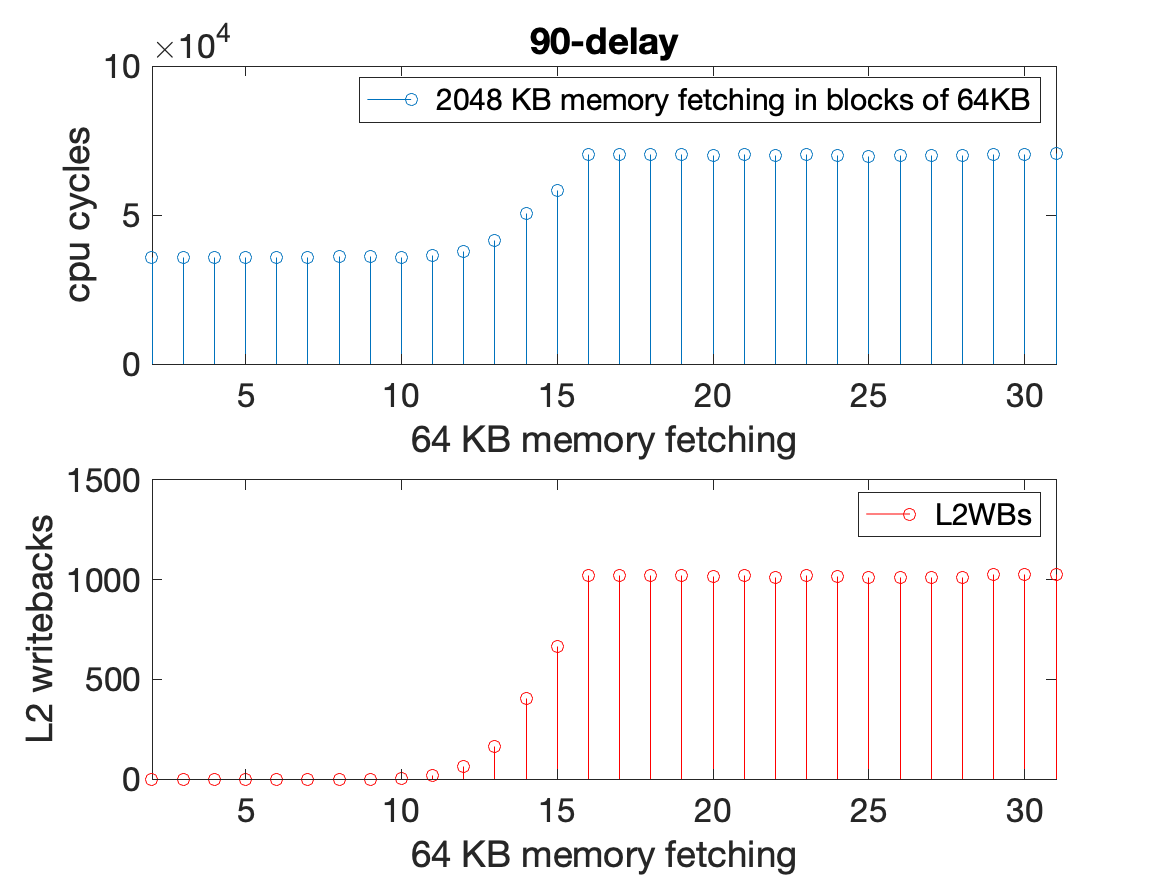} \
\caption{2048 KB memory fetching in several blocks of 64KB when the MBA 90-delay is enabled\label{fig:fineg_l2_wb_90}}
\end{figure}

We include the two L2 traffic patterns as a factor motivating this choice in Fig.\ref{fig:fineg_l2_wb_00} and Fig.\ref{fig:fineg_l2_wb_90} where we fetch 2 MB of dynamic memory in several blocks of 64 KB when the MBA is disabled (0-delay) and enabled with 90-delay. 
While the above plot represents the latency (CPU cycles) for each block, the graph below shows the respective L2WBs for each fetched block. The number of L2WBs grows as the L2 cache fills its capacity (L2 size is 1 MB). While initially, no L2WBs are observed (when L2 capacity is available), subsequently, when the L2 becomes full, each memory read will sure correspond to one L2WB (so 1024 memory reads of 64 bytes (64KB) generate 1024 L2 writebacks). \newline
Looking at the figures, we have two main observations. (i) Without throttling (0-delay), the memory load latency does not change significantly when the L2WBs are present (Fig.\ref{fig:fineg_l2_wb_00})\footnote{Note that our memory read workloads are designed to not cause LLC writebacks (LLCWBs)}. However, (ii) the latency changes when the MBA is enabled (Fig.\ref{fig:fineg_l2_wb_90}) on the two traffic patterns. From the results on CPU cycles, it is evident that the MBA delays the L2WBs requests, doubling the latency of read operations when writebacks are present. 
This might be possible because the controller should introduce a delay towards all outgoing requests on the interconnection network, thus also including reads and writes to the last level cache. \newline
The second observation motivates us to consider L2 traffic pattern as a factor. The $R^{I}$ and $R^{II}$ in Table \ref{tab:doe_onecoreanalysis} represent the read workload respectively in the first (without L2WBs) and the second operational condition (with L2WBs). 
In particular, the read workload is composed of 8192 read requests where each request is an LLC miss ($ readReqs = LLC misses = CAS Reads = 8192$) and hence a memory read operation (CAS R is the command issued on the DRAM channel). The number of 8192 requests (meaning 512KB of memory fetching  because each request retrieves 64 bytes which is the cache line size) has been chosen since it assures us to not have L2WBs in $R^{I}$ ($L2WBs = 0$), given the L2 cache size of 1MB. This is also visible from Fig.\ref{fig:fineg_l2_wb_00}, where after 8 fetchings of 64 KB each, the number of L2WBs is still 0.
On the opposite, in $R^{II}$ the total L2 write-backs must be equal to the total memory requests ($L2 WBs=8192$). This is achieved by performing a pre-read workload to fill the L2 cache.
Note that, considering the results in Fig.\ref{fig:fineg_l2_wb_00} and Fig.\ref{fig:fineg_l2_wb_90}, $R^{I}$ is the worst case workload (it has less limitation), since the frequency (latency) of memory reads is not reduced (delayed) by concurrent L2WBs. 

As a side note, looking at Fig.\ref{fig:fineg_l2_wb_90}, we can observe that the MBA keeps the same limitation for each of the 1024 memory reads. Such workload in software-based approaches would require to set an infeasible regulation period of 10 microseconds, hence proving the better granularity offered by MBA-based approaches.

\subsubsection{Memory write workload}

We have memory writes in a writeback policy when a dirty LLC cache line is evicted.
Hence, since our write workload reproduces the worst case, every request of the workload causes one memory read and one memory write operation, due to the cache line eviction and substitution. 
The $W^{II}$ in  Table \ref{tab:doe_onecoreanalysis} represents this kind of workload.
In particular, the $W^{II}$ workload is composed of 8192 requests where each request is the cause of one memory read, and one memory write ($ writeReqs = LLCmisses = CAS Reads = CAS Writes = 8192$). Only the second operational condition ($II$) is possible for the write workload since an LLC writeback can only be the result of a L2 writeback.

\subsection{Multi-core analysis}
While the single-core analysis gives us information about the memory access limitation when the MBA delay and workload are fixed, the multi-core analysis analyzes the assured memory fetching latency for a critical core when the MBA limits the accesses of the other cores. 
For our evaluation, we deploy a quad-core system where one of them is critical.\newline
In particular, we consider a specific case where the critical core fetches its dynamic memory without MBA limitation, reproducing a workload $R^{I}$. In contrast, the MBA limits the other three cores that perform the most permissive workload $k'$ under limitation resulting by the single-core analysis.
The range of delays considered for the interference cores range from 10 to 90, however we will consider only significant ones $D'$, depending on the result of single-core analysis. The Table \ref{tab:doe_multicoreanalysis} describes the design of experiments. The multi-core analysis requires thirty observations to achieve a 99\% of accuracy.
\subsection{Measurement tool and workload considerations \label{measuremnt_tool}}
We performed experiments in Linux as the high server boot duration (several minutes) makes it time-consuming to do bare-metal development and experimentation. \newline
Our measurement tool captures the latency and several factors (such as LLC writeback, L2 writeback, and CAS commands) of a particular workload. The latency is captured in terms of TSC (Timestamp counter). The fixed TSC frequency is close to the maximum CPU frequency. Hence, we refer to TSC cycles as CPU cycles. We favor the latency rather than bandwidth measurements because the latter requires a regulation period limiting the resolution. \newline
We maximize the interference writing the workload as consecutive memory operations (full-loop unrolling) and choosing a stride between consecutive accesses equal to the cache line size (our cache line size is 64 Bytes). Even though the loop unrolling of memory requests increases the code size, we design our experiments controlling the absence of extra LLC misses. \newline
We reproduce the L2 traffic patterns for a read workload through initial setup operations. To avoid L2 writebacks and to have memory accesses, we flush the dynamic memory before the workload execution, while to reproduce L2 writebacks for the $R^{II}$ workload, we also perform a pre-read workload to fill the L2 cache, as previously mentioned. \newline
Contrarily, for the write workload $W^{II}$, we reproduce the worst-case executing a pre-workload of write-requests filling the entire cache space of dirty and useless cache-lines. In addition to synchronous memory accesses, the previous injected dirty cache lines assure LLC writeback (memory write) whenever an LLC eviction occurs.

\begin{table}
\caption{Design of experiments: multi core analysis \label{tab:doe_multicoreanalysis}}
\begin{center}
\begin{tabular}{ |c| c| }
\hline 
\textbf{Parameters} & \textbf{setting}
\tabularnewline
\hline 
crit. cores no. & $1$
\tabularnewline
\hline
crit. workload & $R^{I}$
\tabularnewline
\hline 
int. cores no. & $3$
\tabularnewline
\hline 
int. workload & the most permissive workload $k' \in K$ under $D'$
\tabularnewline
\hline 
\hline 
\textbf{Factors} & \textbf{values}
\tabularnewline
\hline 
MBA delay& $d_{i}$ for the i-th interference core \\& $d_i\in D'$, $i \in \{1;2;3\}$
\tabularnewline
\hline 
\hline 
\textbf{Metric} & \textbf{description}
\tabularnewline
\hline 
Memory & critical memory fetching latency in terms of CPU cycles\\fetching&
$L_{k';(d_1,d_2,d_3)}$ \\latency &
\tabularnewline
\hline
\end{tabular}
\end{center}
\end{table}

\section{Results from the one-core analysis}
\label{sec:one-core-res}
This section describes the single-core analysis results where the main goal was to estimate the indirect limitation of memory bandwidth when the MBA is enabled.
Below, we represent our results for the three workloads $R^{I}, R^{II}, W^{II}$. 
 
\subsection{Results representation}
$L_{k;d}$ is the execution latency of the workload $k$ when the $d$ MBA delay is active. \newline
Fig.~\ref{fig:boxplot_lat_onecoreanalysis} shows the $L_{k;d}$ boxplots for each experiment of Table \ref{tab:doe_onecoreanalysis} and a mean interpolation for the observations of the same workload and different MBA delays. \newline
Observing the boxplot width, the latency variability is the same (or lower) when the limitation is enabled or not (0-delay). 
Hence, the MBA seems to be not a new source of variability. The main source of that variability is the periodic DRAM refresh cycles.

\subsection{Results interpretation}
\begin{figure} 	
	\centering
	\includegraphics[width=0.47\textwidth]{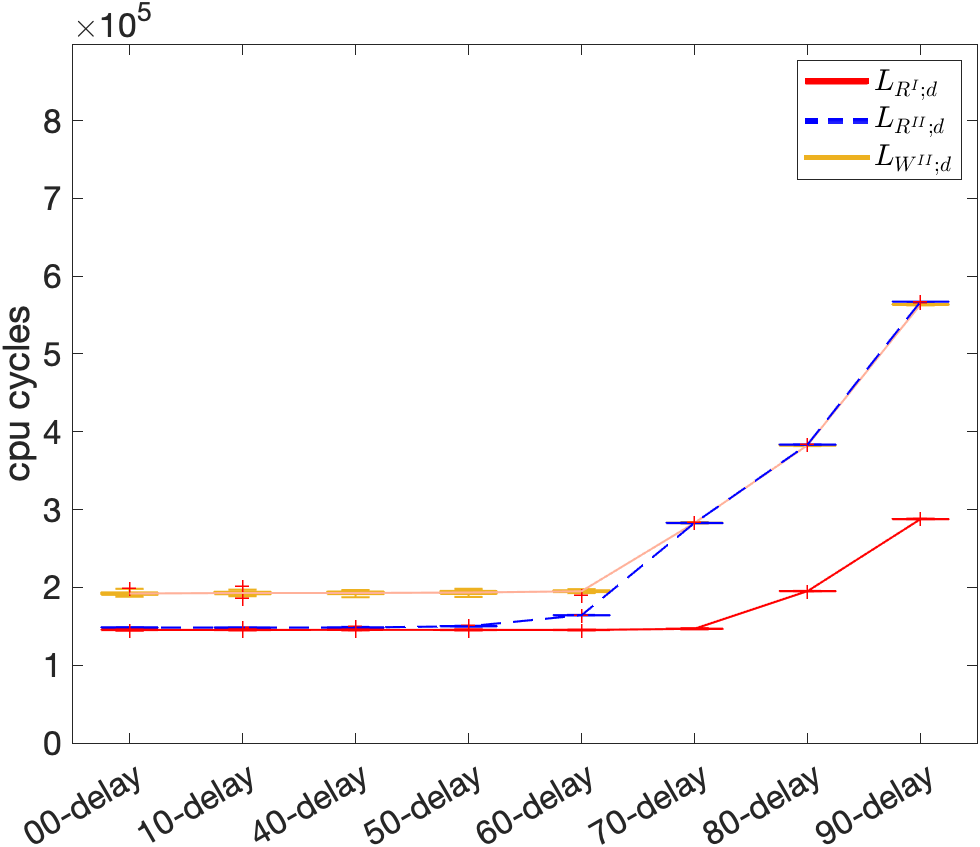}
	\caption{Boxplot latencies for the different workloads and interpolation of the respective mean latencies  \label{fig:boxplot_lat_onecoreanalysis}}
\end{figure}

The Fig.~\ref{fig:boxplot_lat_onecoreanalysis} shows two effects clearly. The first is that the limitations have not an important impact on the relative 0-delay observations until the 70-delay in our setting. The second is that the boxplots of workload $R^{II}$ and $W^{II}$ overlap for high delays, meaning a similar behavior due to L2WBs. 
\newline
To better interpret the results from Fig.~\ref{fig:boxplot_lat_onecoreanalysis} for the 70, 80, and 90 delays, we prefer to go from the latency to bandwidth domain. \newline
Being $T$ the minimum execution latency  $min(L_{R^{I};0})$ to synchronously fetch 8192 cache lines (our chosen reference, as explained earlier),
we calculate the achievable bandwidth (memory operations in $T$) of a generic configuration ${k;d}$, where $k$ is the workload, and $d$ is the delay, as: 
\[
BW_{k;d}=\frac{T*8192}{L_{k;d}}; 
\]
 We report in Table \ref{tab:fineg_max_throug_1op} the maximum observed bandwidth for delays from 60 to 90. \newline
We evaluate also the maximum percentage of synchronous bandwidth as
\[ max(BW_{k;d})\%= \frac{max(BW_{k;d})*100}{8192} \]
Note that the synchronous memory bandwidth could be lower than the achievable memory bandwidth as also memory writes are accounted (i.e., the workload $W^{II}$ achieves $155\%$ of synchronous bandwidth with the less restrictive 60-delay limitation). This is due to the fact that, in the worst case set by our write workload, each write request causes one memory read and one memory write operation, due to the eviction. The "2x" in Table \ref{tab:fineg_max_throug_1op} accounts for this double access regarding memory writes. \newline
Observing the results in the table, we can note that:
\begin{equation}{max(BW_{R^{I};d})>max(BW_{R^{II};d})}\end{equation} 
hence, the less restrictive limitation for a memory read workload is when there are no L2 writebacks, confirming what anticipated in the previous section. We can also note that:
\begin{equation}{
max(BW_{R^{I};70:90})\approx max(BW_{W^{II};70:90})
}
\end{equation} 
that is, the maximum memory accesses in the 1st and in the 2nd operational condition with writes is approximately the same for 70, 80 and 90 delays. Finally, we can note that: 
\begin{equation}
max(BW_{R^{II};70:90})\approx \frac{max(BW_{W^{II};70:90})}{2}
\end{equation} meaning that the memory read requests in the 2nd operational condition for 70,80 and 90 delays are halved due to L2WBs. This regulation choice is pessimistic because L2 writebacks does not always cause LLC writebacks and therefore writes to memory.

The results confirm an important hint for systematic MBA evaluation. When testing the memory bandwidth limitation for read operations, it is necessary to run  measurements without L2 writebacks (first operational condition) capturing the most permissive limitation.  
\begin{table}
\caption{The maximum rate and the percentage of synchronous memory bandwidth corresponding to minimum latency (memory operations in $T$). \label{tab:fineg_max_throug_1op}}
\begin{center}
\begin{tabular}{ |c| c| c| c| c| }
\hline
\textbf{Metrics} & \textbf{60-delay}&\textbf{70-dela}y &\textbf{80-delay} & \textbf{90-delay} 
\tabularnewline[1pt]
\hline 
\hline 
$max(BW_{R^{I};d})$ &8192 & 8079 & 6080 & 4105\\[1pt]
$max(BW_{R^{I};d})\%$ &100\% &98\%  &74\%  &50\%
\tabularnewline[1pt]
\hline 
$max(BW_{R^{II};d})$ &7237&4183&3080&2084\\[1pt]
$max(BW_{R^{II};d})\%$ &88\%&51\%&38\%&25\%
\tabularnewline[1pt]
\hline 
$max(BW_{W^{II};d})$ &2x6376&2x4183&2x3091&2x2097\\[1pt]
$max(BW_{W^{II};d})\%$ &155\%&101\%&75\%&51\%
\tabularnewline[1pt]
\hline
\end{tabular}
\end{center}
\end{table}
In addition, we discovered that even though some delays do not contribute to the indirect memory bandwidth limitation in our setting, the delays 70, 80, and 90 have interesting behavior. Indeed, they limit and conserve their limitation even if there are memory writes. In particular, the 80 and 90 delays limit the synchronous memory bandwidth to 75\% and 50\%, respectively. In contrast, the 70 delay only avoids the overload of asynchronous requests due to L2WBs, halving the synchronous bandwidth in the second operational condition. \newline These are important aspects to be considered when dimensioning a system with MBA in order to assure a given memory bandwidth to a critical core. \newline
Along this direction, in the multicore analysis, we choose to evaluate the assured bandwidth when the MBA limits the interference workloads $R^{I}$  only with 80 and 90 delays, covering the significant cases of synchronous bandwidth limitation.

\section{Results from the multi-core analysis}
\label{sec:multi-core-res}
\begin{figure}
\centering
\includegraphics[width=0.5\textwidth]{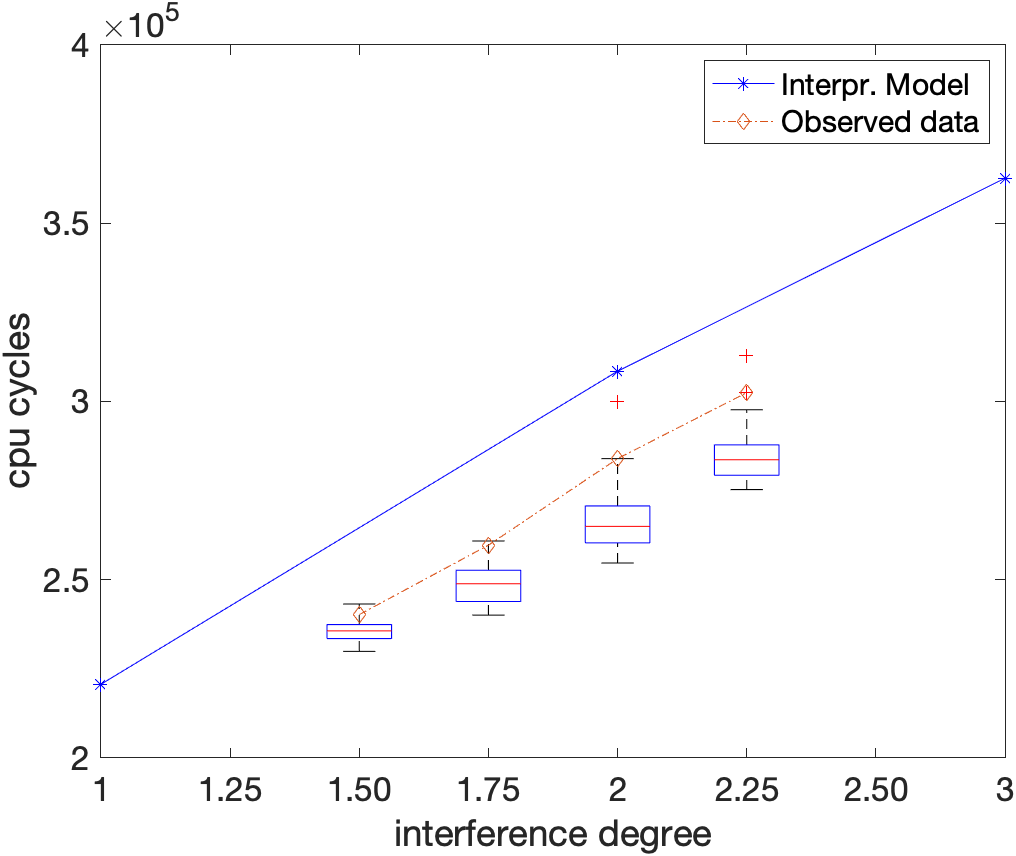}
\caption{Comparison between the critical execution latencies of the interpretative model and those observed as the degree of interference varies
\label{fig:predictting_model}}
\end{figure}

While the single-core analysis focused on the memory bandwidth limitation, we analyze the guaranteed bandwidth of a critical core when the MBA throttles the other cores in the multi-core analysis. \newline
We analyze the delays that significantly limit the synchronous memory bandwidth: 80 and 90 delays reduce synchronous bandwidth by 75 and 50 percent, respectively. In addition, we design an interpretative model to explain the observed data. \newline
We choose $R^{I}$ as interfering workload, as motivated by the single-core analysis results. In the experiments, we deploy a quad-core system where all four cores start with empty private caches. In this way, the interference cores reproduce the $R^{I}$ workload, while the critical core emulates a critical task that fills its private caches. In particular, the critical task fetches 512KB, while the interference cores 768KB, to be sure to last longer than the critical task. \newline
In Table \ref{tab:multicore_combinations_predicted}, we associate an interference degree $id$ to each experiment combination considering the limitation calculated in the single-core analysis.\footnote{Note that we are assuming the combinations position-independent (i.e., if we limit core 1 memory accesses, we suppose to be the same of limiting the core 2).} An interference degree, for us, represents the synchronous memory accesses that we would execute without interference. If that load equals the number of requests performed by a single-core with a maximum rate, then the interference degree is equivalent to one. \newline
The interference degree considers only the limitation of the single-core analysis and not the observed interference. Hence, for instance, the third combination of Table \ref{tab:multicore_combinations_predicted}, where two cores interfere with 80-delays (75\% of synch. bandwidth) and one core with 90-delay (50\% of synch. bandwidth), has the same interference degree of two cores that interfere without limitation.
In Fig. \ref{fig:predictting_model}, we report the boxplots of our critical latency observations for each combination of interference degrees in Table \ref{tab:multicore_combinations_predicted}, and we interpolate the 95-percentile. As expected, the lower the interference degree (achieved with high MBA delays) the lower the impact on the latency of the critical workload. 
\newline
To better interpret this result, we also report the interpolation of the 95-percentile of the critical core latency when having exactly 1, 2 or 3 interfering cores with 0-delay, that is, no MBA limitation (blue line of Fig.~\ref{fig:predictting_model}).

Comparing the two 95-percentile interpolations, we can see how the MBA is effective at limiting the interference. For instance, when the interference degree is set to 2 (with 1 interfering core with 90-delay and 2 interfering cores with 80-delay) the MBA is able to reduce the interference of 3 cores to at most the one we would have with 2 interference cores running without limitation. The blue line can thus be seen as the worst-case memory access latency we can achieve, when varying the interference degree, which is useful to establish a predictable behavior of the critical workload when run in a cloud system. This worst-case line is also easily achievable, from a practical point of view, by running interfering workloads with an increasing number of cores with no MBA. \newline
Hence, our experiments show that the MBA can provide several safety upper bounds for the memory fetching phase of a critical task in a mixed-criticality Cloud infrastructure.Such bounds can be used in the WCET analysis of real-time tasks, without requiring invasive interventions to task code or execution.
In particular, this solution doesn't require to modify the program of non-critical task to respect execution models (contrarily to \cite{Pellizzoni2011APE}, \cite{Boniol2012DeterministicEM}) and it preserves private resource utilization limiting only the access versus the shared context (contrarily to software solutions such as Memguard \cite{Yun2013MemGuardMB} which halt the core execution).


\begin{table}
\caption{Critical core latency variability when the three interference cores are throttled by 80 or 90 delay\label{tab:multicore_combinations_predicted}}
\begin{center}
\begin{tabular}{ |c| c| c| c| c| }
\hline 
\textbf{ \#80d} & \textbf{\#90d }&\textbf{interference degree $id$ \textbf{calculation}} & \textbf{resulting $id$ }
\tabularnewline
\hline 
\hline 
0&3 & $(0*(3/4)+3*(1/2)$ & $1.5$
\tabularnewline
\hline
1&2 & $(1*(3/4)+2*(1/2)$ & $1.75$
\tabularnewline
\hline
2&1 & $(2*(3/4)+1*(1/2)$ & $2$
\tabularnewline
\hline
3&0 & $(3*(3/4)+0*(1/2)$ & $2.25$ 
\tabularnewline
\hline
\end{tabular}
\end{center}
\end{table}

\section{Conclusions and future work}
\label{sec:concl}
MBA is an interesting and reliable hardware feature for resource limitation in real-time systems. It can be used to regulate the access contention by limiting the arrival rate of the single physical cores. 
In this paper, we provide a systematic approach to estimate the indirect limitation of the MBA using three main workloads: memory reads with no L2 writebacks, memory reads with L2 writebacks and memory reads with L2 writebacks and memory writes.
With experiments conducted on real hardware we discover that a good approach to measure the limitation in the worst-case is to run measurements without L2 writebacks. In addition, we found that  only the Intel's 70, 80, and 90 delays values lead to significant limitations un our setting. Such limitation for the maximum synchronous bandwidth continues to be valid when there are memory writes, differently from state-of-the-art software-based solutions such as Memguard \cite{Yun2013MemGuardMB}. \newline
Our evaluation in the multi-core analysis showed that the MBA is able to provide memory bandwidth isolation for a critical task under heavy memory-intensive workloads on multi-core platforms. In addition, the MBA enables short regulation periods that would be hardly achievable with software-based methods.  Finally, the MBA hardware limitation does not require to stop the execution of interfering cores to prevent memory accesses, and it does not require to execute the periodic regulation subroutine, improving the overall efficiency. Although results are achieved for a specific processor, we believe the proposed method is general enough to be applicable on different processor models.\newline
In the future, we plan to design a MBA-based approach to regulate the memory demand based on the current status of the memory controller and requirements of critical tasks, by dynamically assigning the required delay to interference cores, approaching an efficient regulation algorithm.

\section*{Acknowledgements}
This work has been supported by the project COSMIC of UNINA DIETI.

\bibliography{bibliography}
\bibliographystyle{plain}

\end{document}